\documentclass[a4paper,fleqn,usenatbib]{mnras}

\usepackage{graphicx}	
\usepackage{amsmath}	
\usepackage{amssymb}	
\usepackage{hyperref}
\usepackage{footnote}
\usepackage{tablefootnote}
\usepackage{mathptmx}
\usepackage{txfonts}

\usepackage[T1]{fontenc}
\usepackage{aecompl}

\title[ ]{Tracing star formation with non-thermal radio emission}

\author[J. Schober et al.]{
Jennifer Schober,$^{1}$\thanks{E-mail: jschober@nordita.org}
D.~R.~G.~Schleicher,$^{2}$
R.~S.~Klessen$^{3,4}$
\\
$^{1}$Nordita, KTH Royal Institute of Technology and Stockholm University, Roslagstullsbacken 23, 10691 Stockholm, Sweden\\
$^{2}$Departamento de Astronom\'ia, Facultad Ciencias F\'isicas y Matem\'aticas, Universidad de Concepci\'on, \\ Av. Esteban Iturra s/n Barrio Universitario, Casilla 160-C, Concepci\'on, Chile\\
$^{3}$Universit\"at Heidelberg, Zentrum f\"ur Astronomie, Institut f\"ur Theoretische Astrophysik, Albert-Ueberle-Strasse~2, D-69120 Heidelberg, \\Germany\\
$^{4}$Universit\"at Heidelberg, Interdisziplin\"ares Zentrum f\"ur Wissenschaftliches Rechnen, Im Neuenheimer Feld~205, D-69120 Heidelberg, \\Germany
}

\date{Accepted XXX. Received YYY; in original form ZZZ}

\pubyear{2016}

\begin{document}
\label{firstpage}
\pagerange{\pageref{firstpage}--\pageref{lastpage}}
\maketitle

\begin{abstract}
A key for understanding the evolution of galaxies and in particular their star formation history will be future ultra-deep radio surveys. While star formation rates (SFRs) are regularly estimated with phenomenological formulas based on the local FIR-radio correlation, we present here a physically motivated model to relate star formation with radio fluxes. Such a relation holds only in frequency ranges where the flux is dominated by synchrotron emission, as this radiation originates from cosmic rays produced in supernova remnants, therefore reflecting recent star formation. At low frequencies synchrotron emission can be absorbed by the free-free mechanism. This suppression becomes stronger with increasing number density of the gas, more precisely of the free electrons. We estimate the critical observing frequency below which radio emission is not tracing the SFR, and use the three well-studied local galaxies M~51, M~82, and Arp~220 as test cases for our model. If the observed galaxy is at high redshift, this critical frequency moves along with other spectral features to lower values in the observing frame. In the absence of systematic evolutionary effects, one would therefore expect that the method can be applied at lower observing frequencies for high redshift observations. However, in case of a strong increase of the typical gas column densities towards high redshift, the increasing free-free absorption may erase the star formation signatures at low frequencies. At high radio frequencies both, free-free emission and the thermal bump, can dominate the spectrum, also limiting the applicability of this method.

\end{abstract}




\section{Introduction}

Theory suggests that the first galaxies have formed at redshifts $z$ between 10 and 15 in atomic cooling halos and developed further through accretion and mergers \citep{BrommYoshida2011,Conselice2014,MadauDickinson2014,SomervilleDave2015}. The evolution of a galaxy is influenced by various physical processes like turbulence \citep[e.g.][]{WiseTurkAbel2008,GreifEtAl2008}, feedback from stars \citep[e.g.][]{CeverinoKlypin2009} and active galactic nuclei \citep[e.g.][]{BowerEtAl2006,HopkinsEtAl2016}, merger and accretion rates \citep[e.g.][]{LotzEtAl2008}, and possibly also magnetic fields \citep{PakmorSpringel2013}. Consequently, theoretical models of galaxy evolution typically depend on a number of free parameters, which can, however, be constrained by ultra-deep observations of modern telescopes. In the \textit{Hubble Ultra Deep Field} galaxies have been detected at redshifts of approximately 8 and above \citep{BouwensEtAl2010,McLureEtAl2010,BouwensEtAl2011} and owing to the gravitational lensing effect, detailed studies are possible for individual high-$z$ galaxies \citep{IvisonEtAl2010}. Several surveys have been performed at radio wavelengths \citep{GarrettEtAl2002,GruppioniEtAl2003,AppletonEtAl2004,JarvisEtAl2010,SargentEtAl2010,BourneEtAl2011} with the goal of investigating correlations with infrared luminosities. These surveys will soon be complemented by the new generation of radio telescopes, which will perform exhaustive surveys of the 'cosmic dawn'. The data sets from the \textit{LOw Frequency ARray}\footnote{http://www.lofar.org/} (LOFAR) and the \textit{Square Kilometer Array}\footnote{https://www.skatelescope.org/} will be extremely important to constrain theoretical models for galaxy evolution. Hence it is crucial to develop theoretical tools for interpreting spectra as well as fluxes at single frequencies. While local calibrations between the star formation rate and the radio flux have been established in the literature \citep{Condon1992,Bell2003,MurphyEtAl2011}, it is less clear whether these can be extrapolated to very low frequencies and the high-redshift Universe. In this paper we present a physical model for the non-thermal radio luminosity, on the basis of which we are able to discuss limitations regarding its reliability as a SFR tracer.  \\
The origin of non-thermal radio emission from star-forming galaxies are synchrotron losses of highly energetic charged particles, so-called cosmic rays, that spiral around magnetic field lines \citep{BlumenthalGould1970, Longair2011}. The main source of galactic cosmic rays are likely shock fronts in supernova remnants, where charged particles undergo first-order Fermi acceleration \citep{Bell1978a,Bell1978b,Drury1983,Schlickeiser2002}. As the rate of supernovae is related to the rate at which stars form, one can expect a connection between galactic synchrotron emission and the star formation rate (SFR) \citep{Condon1992}. This coupling is reflected in the FIR-radio correlation which has been observed in the local Universe \citep{NiklasBeck1997,YunEtAl2001} and seems to hold up to $z\approx3$ \citep{SeymourEtAl2009,JarvisEtAl2010,SargentEtAl2010,BourneEtAl2011,MagnelliEtAl2015,PannellaEtAl2015}. The physical interpretation of this correlation is based on star formation, which is related to cosmic rays, and thus synchrotron emission, as well as to FIR emission, which origins from dust heated by stellar UV radiation \citep{Bell2003,GrovesEtAl2003,LackiEtAl2010a,LackiEtAl2010b,SchleicherBeck2016}. 
It has been suggested that the FIR-radio correlation can, however, break down at high redshifts due to increasing energy losses of cosmic rays in the stronger cosmic microwave background and at higher gas densities \citep{SchleicherBeck2013,SchoberEtAl2016}. Additionally, the ratio of infrared to radio fluxes has been found to be lower in metal-poor galaxies, which can be interpreted as a result of a lower dust amount and higher dust temperatures and hence should not affect the relation between the SFR and the radio emission \citep{QiuEtAl2017}. 
\\ 
Besides synchrotron emission, free-free emission contributes to the radio spectrum, especially at frequencies between 10 and 100 GHz. The recent KINGFISH survey of nearby galaxies has revealed that this thermal component makes up more than 20 percent of the mid radio flux on average and hence can significantly modify the overall slope of the radio spectrum \citep{TabatabaeiEtAl2016}. When using non-thermal radio emission as a tracer for the SFR, one needs to be particularly careful at low frequencies, where synchrotron emission gets exponentially suppressed by free-free absorption. This is a consequence of the frequency dependency of the optical depth $\tau_\mathrm{ff}$. In fact, below a critical frequency $\nu_\mathrm{crit}$ $\tau_\mathrm{ff}$ becomes larger than one and hence the medium becomes optically thick \citep{SchoberEtAl2016}. As a result there is no correlation of non-thermal radio emission and the SFR below $\nu_\mathrm{crit}$. This frequency depends strongly on the gas density and the ionization degree in the galaxy. With LOFAR surveys are planned between the key frequencies 15 and 200 MHz where free-free absorption can play a crucial role. SKA will be operating at higher frequencies. Surveys with the SKA are planned at several GHz where the method presented in this paper should be applicable.
\\
One goal of the future deep surveys is to measure the typical galactic SFR as a function of redshift. Several tracers of the star formation rate have been established in the literature all across the electromagnetic spectrum \citep{HaoEtAl2011,MurphyEtAl2011,KennicuttEvans2012}. Most common are the integrated 8-1000 $\mu$m flux, the 60 $\mu$m flux, the H$\alpha$ flux, and the UV flux. However, in surveys with the SKA sources will be detected for which only radio fluxes are available, making a radio calibration of the SFR necessary. Young galaxies typically have higher SFRs \citep{MadauEtAl1998} and higher mean gas densities \citep{DaddiEtAl2005,WilliamsEtAl2014,BelliEtAl2014}. With the free-free processes being sensitive to the gas density, we expect much stronger absorption in the early Universe. On the other hand, the critical frequency below which synchrotron emission is absorbed shifts to lower values in the observed frame as redshift increases. Hence, for galaxies where this method is unsuitable in the local Universe at a fixed observing frequency, a determination of the star formation rate may be possible for similar galaxies at high redshift.\\
In this paper we present an analytical model for the radio emission of galaxies. The description of cosmic rays includes a source term related to the supernova rate and several energy loss channels. Free-free emission and absorption depend strongly on the optical depth which is frequency dependent and a function of gas density and the ionization degree. We use our model to derive $\nu_\mathrm{crit}$ below which synchrotron emission is suppressed for different types of star-forming galaxies. The correlation between the radio luminosity at various fixed frequencies and the SFR is calculated and explored for a large parameter space. Finally, we employ our model to high redshifts and then draw our conclusions.


\section{Physical model for non-thermal radio emission}

\subsection{Properties of cosmic ray electrons}

Supernova remnants are most likely the birthplace of galactic cosmic rays, where charged particles gain relativistic energies through first order Fermi acceleration in shock fronts \citep{Bell1978a,Bell1978b,Drury1983,Schlickeiser2002}. The result of this process is a power law distribution in energy that is observed over more than ten orders of magnitude \citep{Hillas2006}. We concentrate here on the cosmic ray electrons, as they are responsibly for the largest contribution to the synchrotron emission due to their low mass. Shock acceleration leads to the following injected energy of cosmic ray electrons $Q_e$ as a function of the Lorentz factor $\gamma$
\begin{equation}
  Q_e(\gamma) = Q_{e,0}~\gamma^{-\chi}.
\label{eq_Qe}
\end{equation}
The power law index $\chi$ has a typical value between $2.1$ and $2.3$ \citep{BogdanVolk1983} and we choose a value of $\chi=2.2$ for this study. \\
In addition to primary $e^\pm$ cosmic rays from supernovae, secondaries are produced from cosmic ray protons that decay into pions which in turn decay into $e^\pm$. For modelling the spectral energy distribution of $e^\pm$ we follow the work of \citet{LackiBeck2013} that has also been described in \citet{SchoberEtAl2016}. Taking both contributions into account, the normalization of the spectrum (\ref{eq_Qe}) is 
\begin{equation}
  Q_{e,0} = \frac{20^{2-\chi}}{6} \frac{f_\pi}{f_\mathrm{sec}} \left(\frac{m_\mathrm{p}}{m_\mathrm{e}}\right)^{\chi}  ~m_\mathrm{e} c^2~Q_\mathrm{p,0}.
\end{equation}
Here $f_\pi\approx0.4$ is the fraction of protons that decay into pions, $f_\mathrm{sec}\approx0.7$ is the ratio of secondary and total cosmic ray $e^\pm$, $m_\mathrm{p}$ and $m_\mathrm{e}$ are the masses of protons and electrons, and $c$ is the speed of light. The normalization of the proton injection spectrum $Q_\mathrm{p,0}$ can be directly related to the supernova rate $\dot{N}_\mathrm{SN}$ via
\begin{equation}
  Q_{\mathrm{p},0} = \frac{\xi E_\mathrm{SN} \dot{N}_\mathrm{SN} (\chi-2)}{m_\mathrm{p} c^2 ~ \gamma_\mathrm{p,0}^{2-\chi}},
\label{eq_Qp0}
\end{equation}
where $\xi\approx0.1$ is the fraction of the supernova energy $E_\mathrm{SN}\approx10^{51}~\mathrm{erg}$ that is converted into kinetic energy of cosmic rays \citep{Dorfi2000}. The Lorentz factor $\gamma_\mathrm{p,0} = 10^9~\mathrm{eV} / (m_\mathrm{p} c^2) \approx 1$ marks the low energy end of the cosmic ray proton spectrum. \\
When traveling through the interstellar medium, cosmic rays lose energy continuously. The total number $N_e(\gamma)$ of $e^\pm$ can be described as
\begin{equation}
  \frac{\partial N_e(\gamma)}{\partial t} = Q_e(\gamma) + \frac{\mathrm{d}}{\mathrm{d} \gamma} \left[\frac{\gamma}{\tau_e(\gamma)} N_e(\gamma)\right],
\label{DiffLossEq}
\end{equation}
where the energy losses are determined by the cooling timescale $\tau_e$. At steady state we find
\begin{equation}
  N_e(\gamma) = \frac{Q_e(\gamma) \tau_e(\gamma)}{\chi-1}.
\label{DiffLossEq}
\end{equation}
The different energy loss channels are: ionization (ion), bremsstrahlung (brems), inverse Compton scattering (IC), synchrotron emission (synch), and galactic outflows (wind). The individual timescales can be summarized as
\begin{eqnarray}
  \tau_\mathrm{ion} & = & \frac{\gamma}{2.7~c~\sigma_\mathrm{T}~(6.85 + 0.5~\mathrm{ln}\gamma)~n} \label{tau_ion}, \\
  \tau_\mathrm{brems} & = & 3.12 \times10^7~\mathrm{yr}~\left(\frac{n}{\mathrm{cm}^{-3}}\right)^{-1} \label{tau_brems}, \\
  \tau_\mathrm{IC} & = & \frac{3~m_\mathrm{e}~c}{4~\sigma_\mathrm{T}~u_\mathrm{ISRF}~\gamma} \label{tau_IC}, \\
  \tau_\mathrm{synch} & = & \frac{3~m_\mathrm{e}~c}{4~\sigma_\mathrm{T}~u_\mathrm{B}~\gamma} \label{tau_synch}, \\
  \tau_\mathrm{wind} & = & \frac{H}{\mathrm{v}_\mathrm{wind}}.  \label{tau_wind}  
\end{eqnarray}
and result in the total cooling timescale
\begin{equation}
  \tau_\mathrm{e} = \left(\tau_\mathrm{ion}^{-1} + \tau_\mathrm{brems}^{-1} + \tau_\mathrm{IC}^{-1} + \tau_\mathrm{synch}^{-1} + \tau_\mathrm{wind}^{-1}\right)^{-1}.
\label{te}
\end{equation}
Here $\sigma_\mathrm{T}\approx 6.65\times10^{-25}~\mathrm{cm}^2$ is the Thomson cross section, $u_B=B^2/(8\pi)$ is the energy density of the magnetic field $B$, $u_\mathrm{ISRF}$ is the energy density of the interstellar radiation field (ISRF), $n$ is a gas density, $H$ is the galactic scale height, $\mathrm{v}_\mathrm{wind}$ the velocity of the galactic wind $\gamma$ is the $e^\pm$ Lorentz factor.

\subsection{Synchrotron emission in a galactic magnetic field}

In the presence of a magnetic field, cosmic rays perform spiral motions and hence are constantly accelerated. A single electron with a Lorentz factor $\gamma$ results in the spectral power
\begin{eqnarray}
  L_{\mathrm{synch},\nu,\gamma}(\nu, \gamma) = \frac{\sqrt{3}~e^3~ B}{m_e c^2}~\frac{\nu}{\nu_\mathrm{c}(\gamma)}~ \int_{\nu/\nu_\mathrm{c}(\gamma)}^\infty K_{5/3}(x)~\mathrm{d}x, \nonumber \\
\label{eq_singlesynch}
\end{eqnarray}
where $K_{5/3}(x)$ is the modified Bessel function of second kind and $\nu_\mathrm{c}(\gamma) = 3 \gamma^2~e~B/(4\pi~c~m_\mathrm{e})$ \citep[see e.g.~the review by][]{BlumenthalGould1970}.\\
To find the synchrotron emission $L_{\nu,\mathrm{synch}}$ produced from the full population of cosmic rays one has to integrate over the cosmic ray distribution, i.e. 
\begin{eqnarray}
  L_{\mathrm{synch},\nu}(\nu) & = & \int_{\gamma_{\mathrm{e},0}}^\infty L_{\mathrm{synch},\nu,\gamma}(\nu, \gamma) N_\mathrm{e}(\gamma) ~\mathrm{d}\gamma \nonumber \\
  & & \times \int N(\alpha) (\mathrm{sin}(\alpha))^{(\chi+1)/2}~\mathrm{d}\Omega_\alpha.
\label{eq_Lnu}
\end{eqnarray}
The last integral over the pitch angle $\Omega_\alpha$ is roughly 8.9 for a cosmic ray spectrum with a slope of $\chi=2.2$ and the lower limit of the $\gamma$ integration is $\gamma_{\mathrm{e},0} = 10^7~\mathrm{eV}/(m_\mathrm{e} c^2) \approx 20$. \\
We note that the synchrotron luminosity $L_{\mathrm{synch},\nu}$ is directly proportional to the supernova rate $\dot{N}_\mathrm{SN}$, which determines the total number of cosmic rays (see equation \ref{eq_Qp0}). As the supernova rate is correlated with the star formation rate, synchrotron emission can be used to estimate a galaxy's SFR.

\begin{table*}
\begin{center}
     \begin{tabular}{lllll}
      \hline  \hline  
      parameter 					& abbreviation 					& normal star-forming galaxy	& starburst core \\
      ~ 						& ~						& (Milky Way)			& (M 82) \\
      \hline
      magnetic field strength $[\mu\mathrm{G}]$		& $B~(B_0)$ 					& $1-20~(10)$			&  $10-100~(50)$	 \\
      star formation rate $[M_\odot~\mathrm{yr}^{-1}]$	& $\dot{M}_\star~(\dot{M}_{\star,0})$ 		& $0.1-10~(2)$ 			&  $10-500~(10)$ \\
      gas density $[\mathrm{cm}^{-3}]$ 			& $n~(n_0)$ 					& $0.1-10~(2)$ 		  	&  $10-1000~(300)$ \\
      intrinsic ISRF $[\mathrm{erg}~\mathrm{cm}^{-3}]$	& $u_\mathrm{int}~(u_\mathrm{int,0})$ 		& $10^{-13}-10^{-11}~(10^{-12})$&  $10^{-10}-10^{-8}~(10^{-9})$	 \\
      scale height $[\mathrm{pc}]$			& $H~(H_0)$ 					& $250-1000~(500)$		&  $100-400~(200)$	 \\
      wind velocity $[\mathrm{km}~\mathrm{s}^{-1}]$ 	& $\mathrm{v}_\mathrm{wind}~(\mathrm{v}_\mathrm{wind,0})$ 		& $1-100~(50)$			&  $10-500~(230)$	 \\
      electron temperature $[\mathrm{K}]$ 		& $T_e~(T_{e,0})$				& $5\times10^3-1.5\times10^4~(10^4)$  	&  $2.5\times10^3-10^4~(5\times10^3)$ \\ 
      ionization degree					& $f_\mathrm{ion}~(f_\mathrm{ion,0})$ 		& $0.05-0.2~(0.1)$		&  $0.05-0.2~(0.1)$	 \\
      filling factor					& $f_\mathrm{fill}~(f_\mathrm{fill,0})$ 	& $0.05-0.3~(0.2)$		&  $0.05-0.3~(0.2)$	 \\
      \hline  \hline  
    \end{tabular}
\end{center}
\caption{\label{Table_Props}The ranges of the different free parameters which are covered by our model for a normal star-forming galaxy and a starburst galaxy. Fiducial values (indicated by an index "0") for the Milky way and M 82 are given in brackets.}
\end{table*}

\subsection{Free-free emission and absorption}

At low frequencies and high gas densities, the interstellar medium is optically thick and sychnrotron emission, holding the information about the SFR, is absorbed. The optically thick regime is characterized by an optical depth $\tau_\mathrm{ff}$ larger than 1. The value of $\tau_\mathrm{ff}$ depends on the electron temperature $T_\mathrm{e}$, the emission measure $EM$, and the frequency $\nu$: 
\begin{eqnarray}
  \tau_\mathrm{ff}(n,\nu) & = & 0.082\left(\frac{T_\mathrm{e}}{\mathrm{K}}\right)^{-1.35} \left(\frac{EM(n)}{\mathrm{cm^{-6}~pc}}\right)  \nonumber \\
    			  &   & \times \left(\frac{\nu}{10^9~\mathrm{Hz}}\right)^{-2.1},
\label{eq_tauff}
\end{eqnarray}
with
\begin{eqnarray}
  EM(n) \approx n_e(n)^2~ H ~ f_\mathrm{fill}^{-1}.
\label{eq_EM}
\end{eqnarray}
The number density of the free electrons $n_e$ can be related to the gas density via the ionization degree $f_\mathrm{ion}$: $n_e=f_\mathrm{ion} n$. The filling factor $f_\mathrm{fill}$ describes the clumping of the medium \citep{EhleBeck1993,BerkhuijsenEtAl2006,Beck2007}. Hence, the critical frequency $\nu_\mathrm{crit}$ at which $\tau_\mathrm{ff}=1$, determining the transition from the optically thin to optically thick regime, is
\begin{eqnarray}
  \frac{\nu_\mathrm{crit}}{10^9~\mathrm{Hz}} & = &  \left(0.082\left(\frac{T_\mathrm{e}}{\mathrm{K}}\right)^{-1.35} \left(\frac{EM(n)}{\mathrm{cm^{-6}~pc}}\right) \right)^{1/2.1}. 
\label{eq_nucrit}
\end{eqnarray}
In addition, we also expect a positive contribution to the radio spectrum from free-free emission. It can be estimated as
\begin{eqnarray}
  L_{\nu,\mathrm{ff}}(\nu) &=&  2~k~T_\mathrm{e}~c^{-2}~\Delta A~(1 - \mathrm{e}^{-\tau_\mathrm{ff}})~\nu^2.
\label{eq_ff}
\end{eqnarray}
The parameter $\Delta A$ is the surface area of the galaxy which depends of course on the galaxy's size as well as on its orientation along the line of sight. With a scaling proportional to $\nu^2$ the free-free emission affects the radio spectrum mostly at high frequencies. \\
The total spectral radio emission including both, synchrotron emission and free-free effects, is then given as
\begin{eqnarray}
  L_{\nu}(\nu) = L_{\nu,\mathrm{synch}}(\nu)\mathrm{e}^{-\tau_\mathrm{ff}(\nu)}  + L_{\nu,\mathrm{ff}}(\nu).
\label{eq_Lnu}
\end{eqnarray}
The luminosity at a fixed frequency $\nu_0$ can be estimated as $\nu_0 L_{\nu}(\nu_0)$.\\
Equation (\ref{eq_Lnu}) is a function of the SFR, if the synchrotron flux dominates the spectrum at a given frequency. The critical frequency (\ref{eq_nucrit}) gives a lower limit for the frequency regime where this method can be applied. Otherwise, for observations at $\nu\lesssim\nu_\mathrm{crit}$, our method can only be used to estimate upper limits of the SFR. Additionally, if $\nu_\mathrm{crit}$ is too close to the peak frequency of the blackbody radiation from the dust, i.e.~the thermal bump, the SFR signatures are washed out and our method also provides only an upper limit for the star formation rate.


\section{Radio luminosity in the local Universe}

\subsection{Basic assumptions, fiducial models, and parameter ranges}
Our model for non-thermal radio emission includes several free parameters which vary in different individual star-forming galaxies. A list containing the typical range of all the free parameters is presented in Table \ref{Table_Props}. We distinguish here two different cases: a normal star-forming disk galaxy based on the Milky Way and a starburst galaxy based on M 82. The fiducial values of the free parameters are listed in the brackets in Table \ref{Table_Props}. \\
\begin{figure*}[ht]
\centering
  \includegraphics[width=0.85\textwidth]{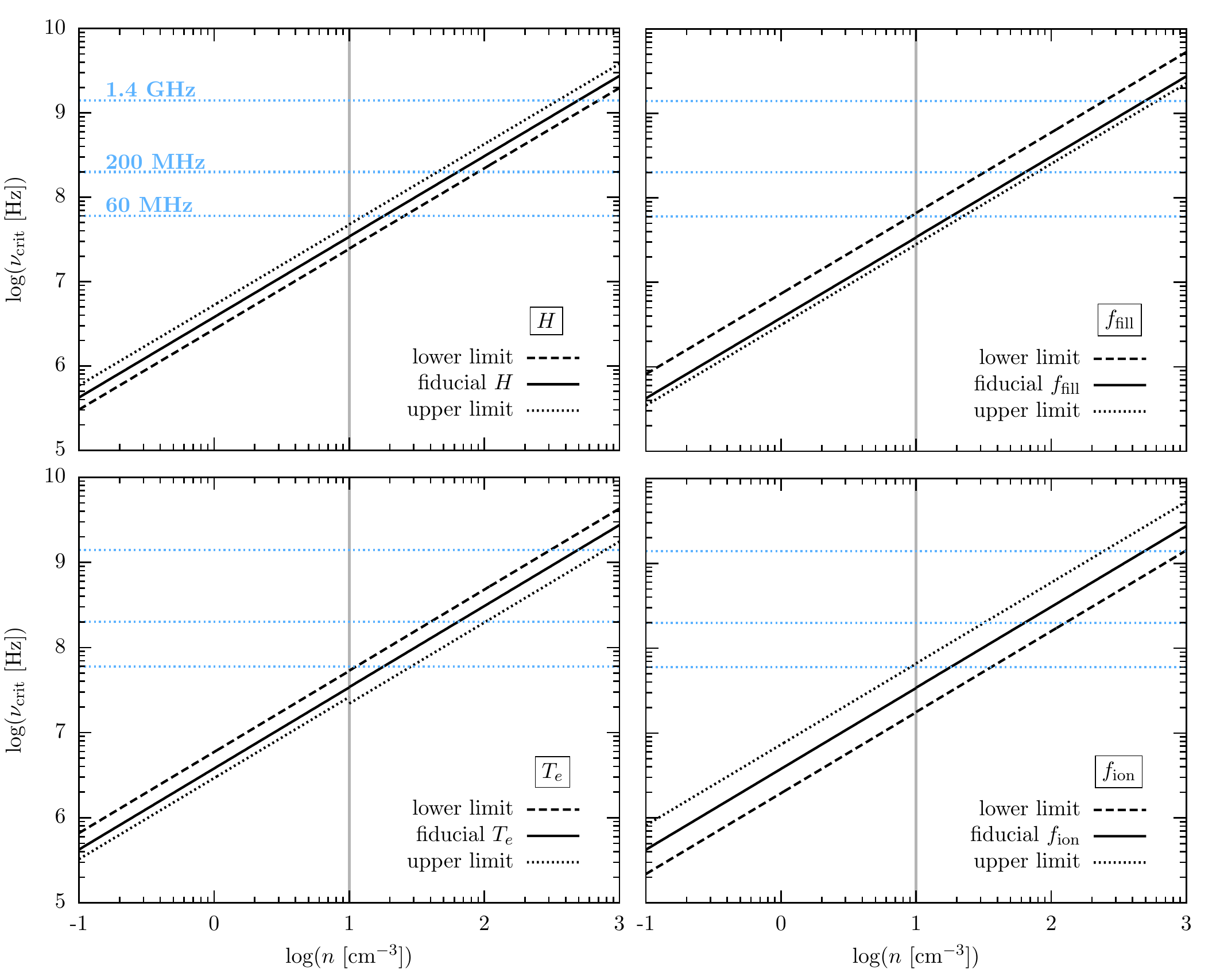}
  \caption{The critical frequency $\nu_\mathrm{crit}$ that is defined via $\tau_\mathrm{ff}(\nu_\mathrm{crit}) \equiv 1$ as a function of gas density $n$. We test the dependency on the scale height $H$ (top left panel), the filling factor $f_\mathrm{fill}$ (top right panel), the electron temperature $T_e$ (lower left panel), and the ionization degree $f_\mathrm{ion}$ (lower right panel). The vertical gray line marks the transition from the normal galaxy model to the starburst model where densities are higher.}
\label{plot_nucrit_n}
\end{figure*}
For determining the synchrotron luminosity the value of the magnetic field strength $B$ is crucial. The total magnetic field observed in spiral galaxies is typically $B=9\pm2~\mu\mathrm{G}$ \citep{Beck2016}, although there are also reports of nearby bright galaxies with $B=17\pm3~\mu\mathrm{G}$ \citep{Fletcher2010}. Gas-rich galaxies with high star formation rates have considerably higher field strengths. \citet{Beck2016} gives a typical value of $B=20-30~\mu\mathrm{G}$. In starburst galaxies values of $50-100~\mu\mathrm{G}$ are observed \citep{ChyzyBeck2004,BeckEtAl2005,HeesenEtAl2011,AdebahrEtAl2013}. \\
The second crucial parameter for the total synchrotron emission is the number of cosmic rays. The cosmic ray injection rate is proportional to the supernova rate $\dot{N}_\mathrm{SN}$ which in turn depends on the SFR $\dot{M}_\star$. The relation between $\dot{N}_\mathrm{SN}$ and $\dot{M}_\star$ is influenced by the choice of the initial mass function (IMF) which determines how many massive stars are forming. Assuming a \citet{Kroupa2002} IMF and a mean mass of stars evolving into supernovae of $\overline{M}_\mathrm{SN} \approx 12.26~M_\odot$ we find
\begin{eqnarray}
  \dot{N}_\mathrm{SN} = 0.156~\frac{\dot{M}_\star}{\overline{M}_\mathrm{SN}}.
\label{eq_NSN}
\end{eqnarray}
For this study we choose a SFR range between 0.1 and 10 $M_\odot\mathrm{yr}^{-1}$ for the normal star-forming galaxies and 10 to $10^3~M_\odot\mathrm{yr}^{-1}$ for the starbursts. \\
Besides the injection, the continuous energy losses determine the steady state number of cosmic rays. The number density of the neutral gas plays here a crucial role. It has been shown that the average midplane density $n$ decreases exponentially with the galactic radius \citep{KalberlaDedes2008}. To simplify the calculation we adopt a single mean density of $n_0=2~\mathrm{cm}^{-3}$ for the Milky Way, but also study a broader range between 0.1 and 10 $\mathrm{cm}^{-3}$ for disk galaxies in general. Densities are much higher in starburst cores. For the case of M 82 \citet{ColbertEtAl1999} report a density of 250 $\mathrm{cm}^{-3}$ if an 3-5 Myr old instantaneous starburst is assumed, while the density can be considerably higher for other scenarios. We use here $n_0 = 300 ~\mathrm{cm}^{-3}$ as a fiducial value, but also consider a larger range of $10-10^3 ~\mathrm{cm}^{-3}$. \\
Losses by inverse Compton scattering are determined by the interstellar radiation field. In Table \ref{Table_Props} we present the values of the intrinsic interstellar radiation field $u_\mathrm{int}$, which refers to the thermal component of the radiation field that is typically related to the SFR. For starburst galaxies $u_\mathrm{int}$ is considerably higher ($u_\mathrm{int}\approx 10^{-9}~\mathrm{erg}~\mathrm{cm}^{-3}$ for M 82) than for example for galaxies with low SFRs \citep[$u_\mathrm{int}\approx 10^{-12}~\mathrm{erg}~\mathrm{cm}^{-3}$ for the Milky Way, see e.g.][]{Draine2011}. In addition, we also include the contribution from the cosmic microwave background (CMB). The latter has an energy density of $u_\mathrm{CMB,0}\approx 4.2\times10^{-13}~\mathrm{erg}~\mathrm{cm}^{-3}$ at redshift $z=0$, but becomes more important in the early Universe. The total interstellar radiation field in our model is calculated as
\begin{equation}
  u_{\mathrm{ISRF}}  =  u_\mathrm{int} + u_\mathrm{CMB,0}(1+z)^4.
\label{uISRFtot}
\end{equation}
The galactic scale height $H$ is important to estimate the losses by outflows (\ref{tau_wind}) and additionally determines the emission measure (\ref{eq_EM}). The thickness of a disk galaxy shows a correlation with its rotational velocity \citep{KregelEtAl2002}. A typical mean value for the Milky Way is 500 pc \citep{KennicuttEvans2012,RixBovy2013}, but we also consider a variation of this value by a factor of 2. For compact starburst cores we choose a value of 200 pc \citep{deCeadelPozoEtAl2009} and again a variation by a factor of 2. 
For the free parameter $\mathrm{v}_\mathrm{wind}$ we refer to a numerical study of a galactic disk by \citet{GirichidisEtAl2016} which shows that the bulk of the outflow occurs at low velocities of $20-40~\mathrm{km~s}^{-1}$. However they also observe a high velocity tail with a few $100~\mathrm{km~s}^{-1}$. We choose for our Milky Way model a value of $v_\mathrm{wind,0} = 50~\mathrm{km~s}^{-1}$ as a reasonable fiducial value. For M 82 an outflow velocity of $230~\mathrm{km~s}^{-1}$ has been observed \citep{WalterEtAl2002}.\\
The free-free optical depth (\ref{eq_tauff}) depends on the electron temperature $T_e$ which can be determined accurately from recombination lines at radio and millimeter wavelengths. \citet{QuirezaEtAl2006} find for the Milky Way a gradient between 4000 to 13000 K with a mean of $T_{e,0}=10^4$ K. Dusty starburst have usually lower excitation temperatures than the Milky Way. For M 82 a value of $T_{e,0}=5000\pm1000$ K has been observed \citep{PuxleyEtAl1989} which we adopt as our fiducial value. In addition the ionization degree $f_\mathrm{ion}$, which determines the number density of free electrons, enters the calculation of $\tau_\mathrm{ff}$. A typical value for the warm interstellar medium is ten percent \citep{Tielens2005}, which we vary by a factor of two. Also the morphological structure of the gas distribution, described by the filling factor, affects the optical depth. A value 0.05 has been considered as a lower limit for $f_\mathrm{fill}$ \citep{EhleBeck1993,BerkhuijsenEtAl2006,Beck2007}. For our fiducial model we use $f_\mathrm{fill}=0.2$, as there are likely additional contributions from ionized components besides to the observed HII regions. The two parameters, $f_\mathrm{fill}$ and $f_\mathrm{ion}$, enter the calculation of $\tau_\mathrm{ff}$ in the combination $f_\mathrm{ion}^2/f_\mathrm{fill}$, and are thus degenerate. Our parameter space ranges from an effective value of approximately 0.008 to 0.8, thus spanning over two orders of magnitude.   \\
For this study the most interesting wavelength regime is where the emission is dominated by synchrotron radiation. We note, however, that the free-free flux scales as $S_\nu\propto\nu^{-0.1}$, while the synchrotron flux scales approximately as $S_\nu\propto\nu^{-0.6}$. Consequently, especially at high frequencies, the free-free emission might contribute significantly to the total observed flux. The equation for the free-free emission (\ref{eq_ff}) includes another free parameter, the surface area of the galaxy $\Delta A$. We employ here a fiducial value of $100~\mathrm{kpc}^{2}$ for the normal disk galaxies and a smaller value of $0.1~\mathrm{kpc}^{2}$ for the compact starbursts. The latter surface area is motivated from the central starburst core of M 82, which has an approximate extension of 200 to 300 pc \citep{deCeadelPozoEtAl2009}.

\begin{figure}
  \includegraphics[width=0.45\textwidth]{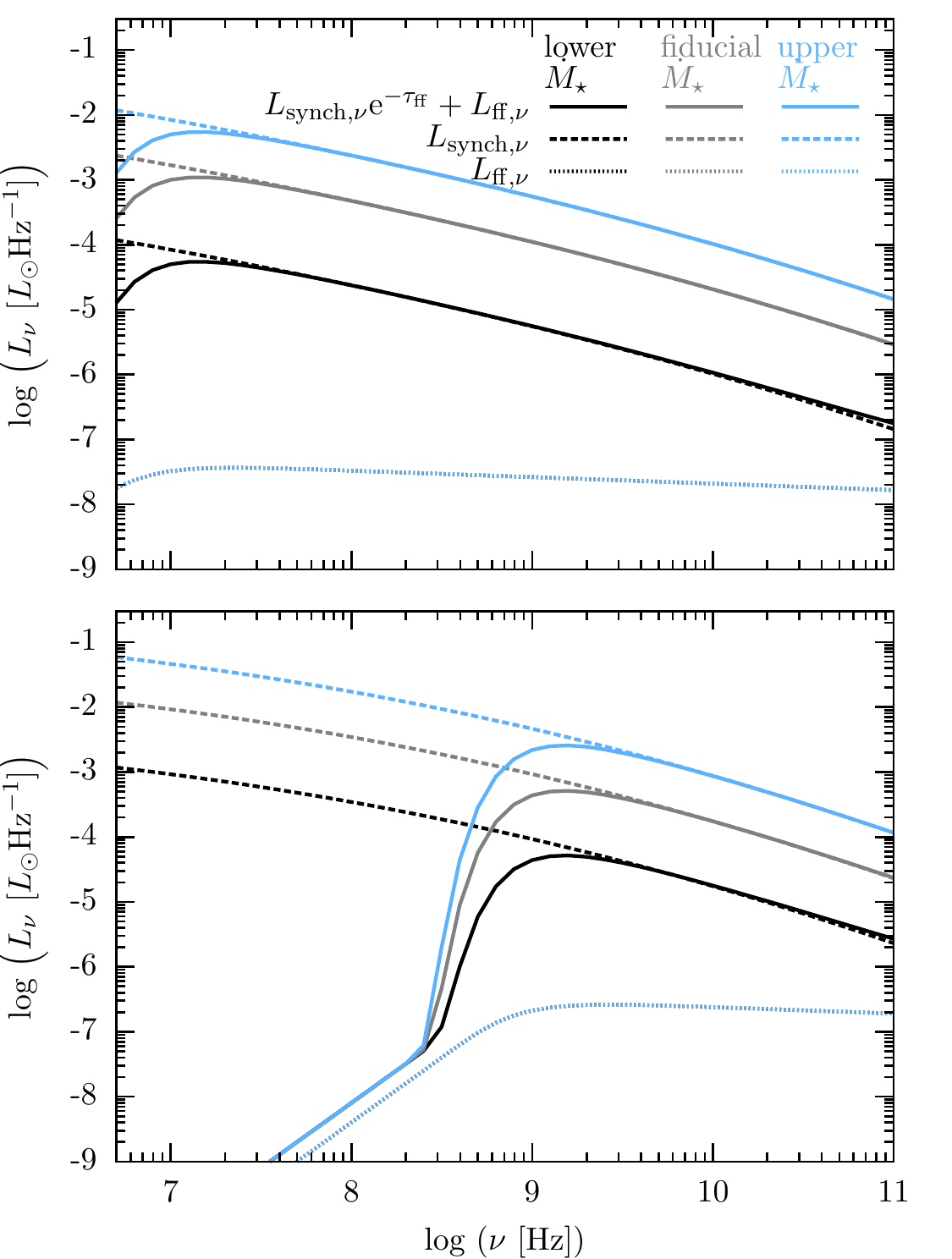}
  \caption{The spectral total radio luminosity $L_{\nu,\mathrm{synch}}(\nu)\mathrm{e}^{-\tau_\mathrm{ff}(\nu)} + L_{\nu,\mathrm{ff}}(\nu)$ (solid lines), the synchrotron component $L_{\nu,\mathrm{synch}}(\nu)$ (dotted lines), and the free-free emission $L_{\nu,\mathrm{ff}}(\nu)$. The fiducial model Milky Way model is shown in the upper panel, the M 82 model in the lower panel. The dependency on the SFR is tested for both cases, with the lower limit of the parameter range given as black curves, the fiducial values as gray curves, and the upper limit as blue curves. The range of SFRs considered for both models is summarized in Table \ref{Table_Props}}.
\label{plot_Lnu_nu}
\end{figure}

\begin{figure}
  \includegraphics[width=0.45\textwidth]{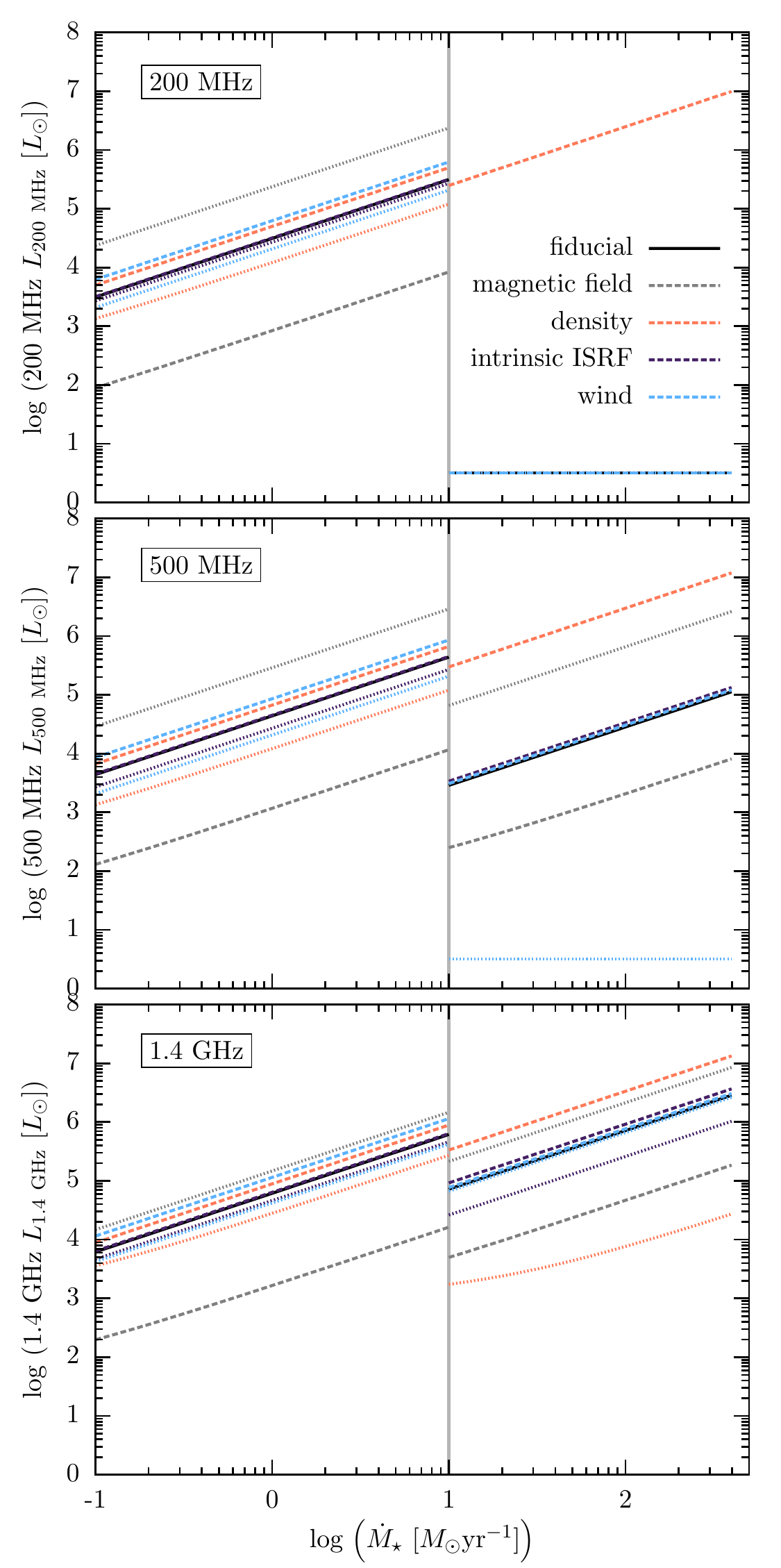}\hfill
  \caption{The total radio luminosity $\nu_0 L_{\nu}(\nu_0)$ as a function of the star formation rate $\dot{M}_\star$. The vertical line at $\dot{M}_\star=10~M_\odot\mathrm{yr}^{-1}$ indicates the transition from our normal disk parameter regime to starburst cores. This sudden transition is rather arbitrary and we expect an overlapping regime in reality, including disk galaxies with higher $\dot{M}_\star$ and starbursts with lower $\dot{M}_\star$. From top to bottom the observing frequencies are $\nu_0=200~\mathrm{MHz}$, $\nu_0=500~\mathrm{MHz}$, and $\nu_0=1.4~\mathrm{GHz}$. The fiducial model is presented in black solid lines. Colored lines refer to the lower limit (dashed lines) and the upper limits (dotted lines) of the parameters listed in the plot legend. The parameter ranges are given in Table \ref{Table_Props}.}
\label{plot_L_SFR}
\end{figure}

\subsection{Using non-thermal radio emission to estimate the SFR}

If synchrotron emission is not suppressed and the radio spectrum is not contaminated by other sources, e.g.~an active galactic nucleus, the non-thermal radio luminosity can be used to estimate the SFR. Strictly speaking, the free-free optical depth $\tau_\mathrm{ff}$ needs to be less than one at the observed frequency, which is the case above the critical frequency $\nu_\mathrm{crit}$ as given in equation (\ref{eq_nucrit}). For our fiducial model based on the Milky Way we find a critical frequency of $\nu_\mathrm{crit}\approx7.3\times10^6$ Hz, while $\nu_\mathrm{crit}\approx8.7\times10^8$ Hz in the fiducial starburst case. We compute $\nu_\mathrm{crit}$ as a function of gas density $n$ and present the result in Figure \ref{plot_nucrit_n}, where we test the dependencies on the different free parameters. The gas density is separated in two regimes. At $n<10~\mathrm{cm}^{-3}$ we use the model for a normal star-forming galaxy, for higher densities the one for the starburst. By the horizontal blue lines we indicate typical frequencies at which observations are performed. The strongest dependency of $\nu_\mathrm{crit}$ is on the density and the ionization degree. Figure \ref{plot_nucrit_n} clearly shows that synchrotron emission is often suppressed in starburst galaxies. Especially at low observed frequencies, e.g.~at 60 MHz, non-thermal radio emission is no tracer for the star formation rate.\\
The latter conclusion is confirmed by Figure \ref{plot_Lnu_nu} where we present the radio spectra for our two fiducial models. We plot the pure synchrotron emission $L_{\mathrm{synch},\nu}$ as dotted lines, pure free-free emission $L_{\mathrm{ff},\nu}$ as dashed lines, and the total radio luminosity as solid lines. In the upper panel the case of a Milky Way like galaxy is presented. Here the bulk of the spectrum is dominated by synchrotron emission and the radio luminosity traces the SFR. In the lower panel we present the starburst case, where $\nu_\mathrm{crit}\approx 870$ MHz. In this case only observations above $\approx 500$ MHz can be employed for estimating the SFR. If the observational frequency is, however, too high, the result might be affected by the contribution of free-free emission. The different colors in Figure \ref{plot_Lnu_nu} refer to different SFRs. As expected from the correlation of the number of cosmic rays, the synchrotron flux decreases with decreasing $\dot{M}_\star$. \\
In Figure \ref{plot_L_SFR} the radio luminosities at different fixed frequencies are plotted as a function of the SFR. With 200 MHz, 500 MHz, and 1.4 GHz, we present typical frequencies used in radio observations. Below 10 $M_\odot~\mathrm{yr}^{-1}$ we use our model for normal disk galaxies, at higher SFRs the starburst model. As a result of the very different parameter values of the two fiducial models, for example a factor of 150 in the gas density (see Table \ref{Table_Props}), there is a sudden jump in the luminosity curve. While this transition at $\dot{M}_\star = 10~M_\odot~\mathrm{yr}^{-1}$ is abrupt in the illustration, we expect an overlapping regime in reality. The results from the fiducial model are shown as black solid lines. For normal galaxies the full parameter range results in a scaling of the radio luminosity with $\dot{M}_\star$. This can be expected, as in this case all the exemplary frequencies are higher than $\nu_\mathrm{crit}$ (see Figure \ref{plot_nucrit_n}). At 200 MHz (upper panel of Figure \ref{plot_L_SFR}) and observations of starbursts, non-thermal radio emission can only be used for extremely low gas densities. For a large range of the parameter space, no correlation with $\dot{M}_\star$ is expected. Observations at 500 MHz (middle panel of Figure \ref{plot_L_SFR}) are close to the critical frequency for starbursts. Here synchrotron emission already becomes suppressed, but a correlation with star formation is still present for a large range of the parameter space. At 1.4 GHz (lower panel of Figure \ref{plot_L_SFR}) the SFR estimate via radio emission is possible for the full range of parameters studied in this paper. \\
Based on these findings we provide the following formulas for our fiducial value of the Milky Way like galaxy at different observing frequencies:
\begin{equation}
 \frac{\dot{M}_\star}{M_\odot~\mathrm{yr}^{-1}} \approx
\begin{cases}
3.20\times10^{-5}~ \dfrac{60~\mathrm{MHz}~ L_{60~\mathrm{MHz}}}{L_\odot} \vspace{0.25cm}\\
2.29\times10^{-5}~ \dfrac{200~\mathrm{MHz}~ L_{200~\mathrm{MHz}}}{L_\odot} \vspace{0.25cm}\\
1.63\times10^{-5}~ \dfrac{1.4~\mathrm{GHz}~ L_{1.4~\mathrm{GHz}}}{L_\odot} .
\end{cases}
\label{eq_correlationfitMW}
\end{equation}
For the fiducial model of starburst cores, we only find a correlation between the SFR and the non-thermal radio luminosity for high frequencies. The M 82 case yields the following formula:
\begin{equation}
 \frac{\dot{M}_\star}{M_\odot~\mathrm{yr}^{-1}} \approx
1.39\times10^{-4} \frac{1.4~\mathrm{GHz}~ L_{1.4~\mathrm{GHz}}}{L_\odot}.
\label{eq_correlationfitM82}
\end{equation}
We stress that these formulas are derived for our two fiducial models, the Milky Way disk and the core of M 82. Based on the uncertainties in observations of the individual input parameters and deviating properties of other galaxies, these formulas give only order of magnitude estimates of the star formation rate. A more accurate calculation of the SFR is possible if some of the free parameters, like the density and the magnetic field strength, are known and inserted into the full model, e.g.~solving equation (\ref{eq_Lnu}) for $\dot{M}_\star$.
\\
The fiducial formulas (\ref{eq_correlationfitMW}) and (\ref{eq_correlationfitM82}) from our model can be compared the SFR calibrator reported in \citet{MurphyEtAl2011}. These authors estimate the relation between the SFR and the integrated IR flux using Starburst99 \citep{LeithererEtAl1999}. By combining this result with the empirical FIR-radio correlation they find the following correlation:
\begin{equation}
 \frac{\dot{M}_\star}{M_\odot~\mathrm{yr}^{-1}} \approx
1.74\times10^{-4} \frac{1.4~\mathrm{GHz}~ L_{1.4~\mathrm{GHz}}}{L_\odot}.
\label{eq_correlationfitMurphy}
\end{equation}
The proportionality factor is close to the one found for our fiducial starburst case.

\subsection{Application to local galaxies}

\begin{figure} 
  \includegraphics[width=0.45\textwidth]{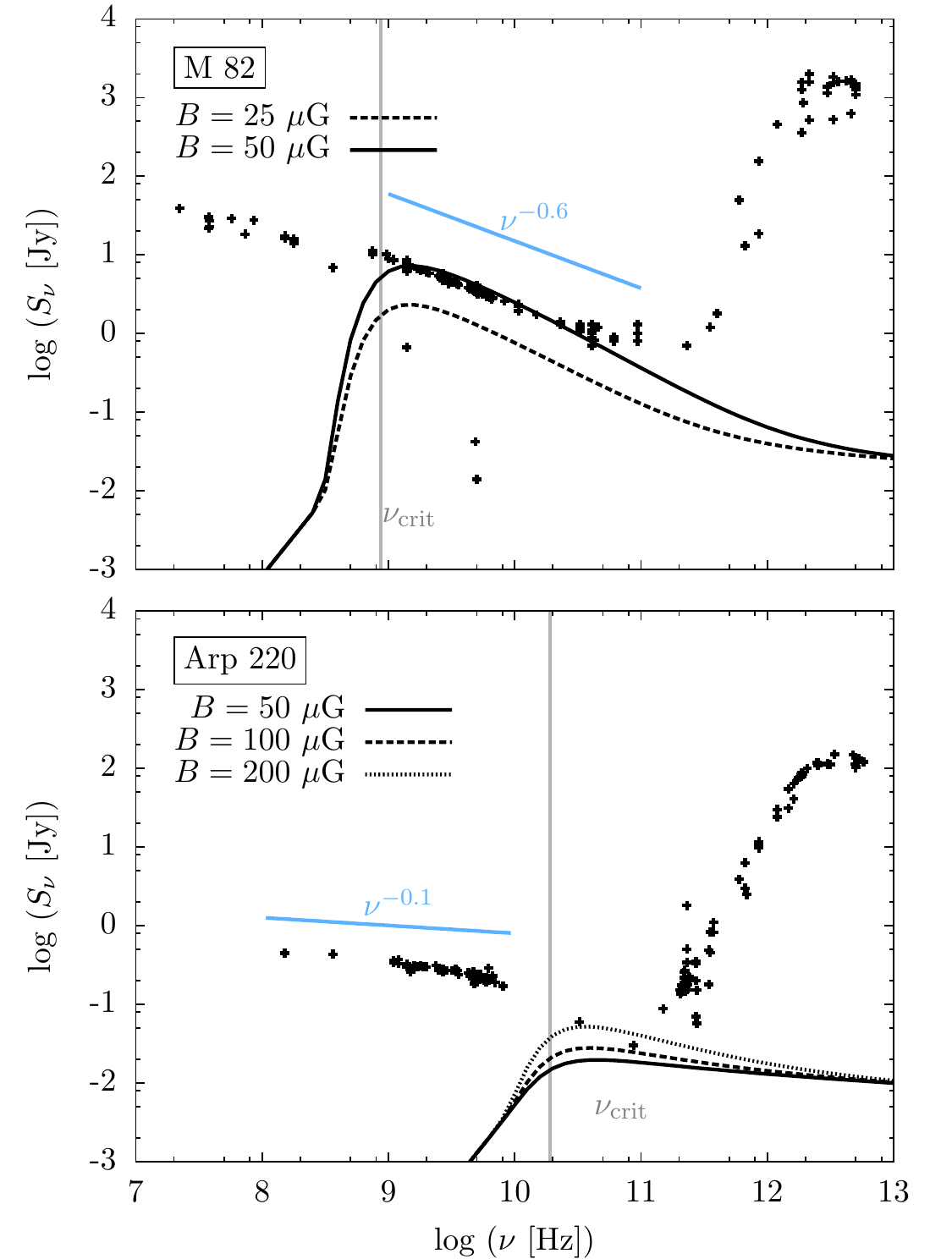}\hfill
  \caption{The observed spectra of two exemplary starburst galaxies: M 82 in the top panel and Arp 220 in the lower panel. The black dots show the photometric data from the NED. In the radio regime we indicate the scaling that results from free-free emission ($S_\nu\propto \nu^{-0.1}$) and the scaling expected from synchrotron emission ($S_\nu\propto \nu^{-(\chi-1)/2}$, i.e.~$S_\nu\propto \nu^{-0.6}$ for our power law exponent of $\chi=2.2$). The critical frequency below which the gas becomes optically thick, $\nu_\mathrm{crit}$ is presented as the vertical gray line. For comparison we present the theoretical curves, where we adopt $\dot{M}_\star=6.8~M_\odot\mathrm{yr}^{-1}$ for M 82 and $\dot{M}_\star=200~M_\odot\mathrm{yr}^{-1}$ for Arp 220. Besides the presenting curves based on the fiducial value of 50 $\mu$G, we also indicate the effect of varying $B$.}
\label{plot_RealSpectra}
\end{figure}

In this section, we apply our model to exemplary local galaxies for which the SFR can also be estimated from traditional tracers. We choose the gas-rich disk galaxy M 51, the core of M 82, which is also our fiducial model for starburst galaxies, and the core region of Arp 220. The input parameters are listed in Table \ref{Table_Examples}. Whenever available, we use estimates from observations, indicated by the references in the table. Often quantities are not well constrained and we give the uncertainty ranges in brackets. Where no further information has been provided in the literature, we consider the errors to be one sigma, leaving the possibility for larger deviations in individual parameters. Where no observational results are reported in literature, we use our fiducial values. The first five parameters listed in Table \ref{Table_Examples}, $n$, $T_\mathrm{e}$, $H$, $f_\mathrm{ion}$, and $f_\mathrm{fill}$, are important to estimate the critical frequency above which we expect significant synchrotron emission. For observing frequencies larger than $\nu_\mathrm{crit}$, we estimate the SFR using public database fluxes. Assuming that the spectrum is dominated by non-thermal radiation, the radio luminosity (\ref{eq_singlesynch}) depends strongly on the magnetic field strength $B$. In fact, a low estimated SFR from our model could be compensated by employing a lower value of $B$. Observations of the magnetic field in galaxies are difficult and rely on certain assumptions, e.g.~an equipartition assumption between the magnetic field and cosmic rays, implying at least a factor of two for the uncertainty of $B$. For M 51, \citet{FletcherEtAl2011} consider a value between 15 and 25 $\mu$G, while a value of 24-98 $\mu$G is discussed for M 82 \citep{AdebahrEtAl2013}.
\\
We test our model for radio observations at 1.4 GHz and 1.5 GHz which are typically above the critical frequency for synchrotron absorption (see e.g.~Figure \ref{plot_nucrit_n}). The critical frequency $\nu_\mathrm{crit}$ can be calculated from equation (\ref{eq_nucrit}). For M 51 and M 82 we estimate critical frequencies of $9.9\times10^6$ Hz and $8.7\times10^8$ Hz, respectively, and thus synchrotron emission is not suppressed by free-free absorption at 1.4 GHz. The density of Arp 220 is roughly four times higher than the one of M 82, and therefore $\nu_\mathrm{crit}$ reaches a value of $1.9\times10^{10}$ Hz, which exceeds an observing frequency of 1.4 GHz. As discussed in the previous paragraph, the parameters determinating $\nu_\mathrm{crit}$ are associated with some uncertainty. We calculate the possible range of $\nu_\mathrm{crit}$ using the lower and upper limits of $n$ and $H$ and report the result in the brackets of Table \ref{Table_Examples}. We find that errors in $\nu_\mathrm{crit}$ can be up to a factor ten due to uncertainties in observational data.\\
The observed radio spectra from the \textit{NASA/IPAC Extragalactic Database}\footnote{https://ned.ipac.caltech.edu/} (NED), which are presented in Figure \ref{plot_RealSpectra}, indicate the suppression of synchrotron emission below $\nu_\mathrm{crit}$. The flux density $S_\nu$ is related to the luminosity via $\nu L_\nu = \nu S_\nu~4 \pi d^2$ where $d$ is the distance of the source. For M 82 the flux density above $\nu_\mathrm{crit}$ scales with approximately $\nu^{-(\chi-1)/2}$ which is $\nu^{-0.6}$ for a scaling of the cosmic ray spectrum with $\chi=2.2$. This is expected for synchrotron emission. At frequencies above a few times $10^{11}$ Hz, the spectrum is dominated by thermal emission. The critical frequency, which is indicated as a vertical gray line, is much higher in the case of Arp 220. Here synchrotron emission is suppressed by free-free absorption. At an observing frequency of 1.4 GHz our model for estimating the SFR should not be used. There are not many photometric data points above $\nu_\mathrm{crit}$ and below the thermal peak for Arp 220, making this galaxy unsuitable for our method. This conclusion is supported by the recent work of \citet{VareniusEtAl2016}, who find that this source only lies on the FIR-radio correlation after correcting for absorption. We note that the observed spectra of both starburst galaxies, M 82 and Arp 220, show significant emission below the respective $\nu_\mathrm{crit}$, even though our model predicts exponential absorption towards lower frequencies. This discrepancy with the observational data is a relic of our simplified galaxy model, in which we assume a uniform gas density. A real galaxy has, however, a gradually decreasing density from the center to the outer regions and is highly inhomogeneous. Consequently, the optical depth becomes typically smaller in the outer galaxy, leading to less absorption of synchrotron and free-free radiation. For Arp 220 the scaling of the spectrum at low frequencies is roughly proportional to $\nu^{-0.1}$. This could be an indication of free-free emission from the halo along the line of sight, in which the gas density is lower and thus the gas is less optically thick. A model for the low frequency spectrum of a starburst galaxy with a focus on M 82 can be found in \citet{Lacki2013}, who describes the HII regions as discrete objects. \\
For direct comparison we overlay our model spectra with the radio data in Figure \ref{plot_RealSpectra}. For M82 we use a star formation rate of $6.8~M_\odot\mathrm{yr}^{-1}$, which is calibrated to the 1.4 GHz observation, while we use a value of $200~M_\odot\mathrm{yr}^{-1}$ for Arp 220. For our fiducial value of $50~\mu$G, the model curve of M 82 agrees well the data above $\nu_\mathrm{crit}$ and below the thermal bump, which dominates at $\nu>10^{11}$ Hz. For the case of a lower magnetic field, like the exemplary $B = 25~\mu$G, which is a better estimate for the outer regions of M 82, the model curve lies lower assuming the same SFR. For Arp 220 we find $\nu_\mathrm{crit}\approx2\times10^{10}$ Hz, and thus do not expect synchrotron emission to dominate a significant range of the radio spectrum. Indeed, our model spectrum, based on $\dot{M}_\star = 200~M_\odot\mathrm{yr}^{-1}$ and $B = 50~\mu$G, appears to have the shape of pure free-free emission and lies slightly below the data points above $\nu_\mathrm{crit}$ and below the thermal bump. Using a higher value of the magnetic field, e.g.~$B = 100-200~\mu$G as presented in the lower panel of Figure \ref{plot_RealSpectra}, is in better agreement with the observational data.  \\
\begin{table*}
\begin{center}
     \begin{tabular}{llllll}
   \hline  \hline  
      ~ 													& M 51 									& M 82 									&  Arp 220	 					  \\
      ~ 													&  									&  									&  							  \\
   \hline		
      input: 													&									&  									&  							  \\			
      $n$ $[\mathrm{cm}^{-3}]$ 											& $5$ ($1 - 5$) \citep{KodaEtAl2009}					& $300$ ($200 - 300$) \citep{ColbertEtAl1999}				& $10^4$ ($>10^3$) \citep{Anantharamaiah2000}  	 	 \\
      $T_\mathrm{e}$ $[\mathrm{K}]$ 										& $10^4$  								& $5\times10^3$								& $7500$ \citep{Anantharamaiah2000} 		  	\\
      $H$ $[\mathrm{pc}]$											& $150$ ($95 - 178$) \citep{HuEtAl2013}					& $200$ ($100 - 600$) \citep{AdebahrEtAl2013}				& $200$ 		 				 \\
      $f_\mathrm{ion}$												& $0.1$   								& $0.1$  								& $0.1$   						  \\
      $f_\mathrm{fill}$									& $0.2$  								& $0.2$                  						& $0.2$  						  \\
      $B$ $[\mu\mathrm{G}]$									& $20$ ($15 - 25$) \citep{FletcherEtAl2011}				& $50$ ($24 - 98$) \citep{AdebahrEtAl2013}				& $50$							  \\
      $u_\mathrm{int}$ $[\mathrm{erg}~\mathrm{cm}^{-3}]$							& $10^{-12}$ 								& $10^{-9}$ 								& $230$  							  \\
      $\mathrm{v}_\mathrm{wind}$ $[\mathrm{km}~\mathrm{s}^{-1}]$ 				& $50$ 									& $230$ \citep{WalterEtAl2002}						& $10^{42}$  							  \\
      $\Delta A$ $[\mathrm{cm}^{2}]$										& $10^{45}$ 								& $10^{42}$								& $10^{42}$  						     	  \\
      $d$ [pc]													& $8.0\times10^6$ (mean from NED) 					& $3.9\times10^6$ (mean from NED)					& $7.7\times10^7$ (mean from NED)  				  \\
   \hline
      observed fluxes:												&  									&  									& 							  \\			
      $S_\mathrm{1.4~\mathrm{GHz}}~[\mathrm{Jy}]$								& $1.4\pm0.1$ \citep{DumasEtAl2011}					& $7.25\pm0.11$ \citep{WilliamsBower2010}				& $0.32\pm0.01$ \citep{WilliamsBower2010}				 \\
      $S_\mathrm{1.5~\mathrm{GHz}}~[\mathrm{Jy}]$								& -			 						& $6.85\pm0.15$ \citep{WilliamsBower2010}				& $0.26\pm0.02$ \citep{WilliamsBower2010} 			\\
      $S_\mathrm{70~\mu\mathrm{m}}~[\mathrm{Jy}]$ 								& $140$	(i)		 						& $1630\pm510$  \citep{DaleEtAl2009}					& $125$ (i)  							\\
      $S_\mathrm{60~\mu\mathrm{m}}~[\mathrm{Jy}]$ 								& $70.3\pm3.0$ \citep{TuffsGabriel2003}					& $1313.5\pm0.6$ \citep{SoiferEtAl1989}					& $103\pm 34$ \citep{SoiferEtAl1989} 			\\
      $S_\mathrm{24~\mu\mathrm{m}}~[\mathrm{Jy}]$ 								& $13$ (i) 								& $325\pm103$ \citep{DaleEtAl2009}					& $5.6\pm0.5$ \citep{BrownEtAl2014}				\\
  \hline
      SFR from our model:											&  									&  									&   							  \\			
      $\nu_\mathrm{crit}~[\mathrm{Hz}]$ (a) 									& $9.9\times10^6~(1.7\times10^6-1.1\times10^7)$  			& $8.7\times10^8~(4.3\times10^8-1.4\times10^9)$ 			& $1.9\times10^{10}~(>2.1\times10^{9})$   			  \\
      $\dot{M}_\star^{1.4~\mathrm{GHz}}~[M_\odot \mathrm{yr}^{-1}]$ (b)						& $0.6~(0.4-0.9)$ 							& $6.8~(2.4-23.1)$ 	    						& -   							  \\
      $\dot{M}_\star^{1.4~\mathrm{GHz}}~[M_\odot \mathrm{yr}^{-1}]$ (c)  					& $0.6$ 								& $6.8$ 	    							& ($118$)   						 \\
      $\dot{M}_\star^{1.5~\mathrm{GHz}}~[M_\odot \mathrm{yr}^{-1}]$ (d) 					& - 									& $6.4~(2.2-21.8)$     	   						& -   							  \\      
  \hline			
      SFR from other tracers:											&  									&  									&    							  \\			
      $\dot{M}_\star^{1.4~\mathrm{GHz}}~[M_\odot \mathrm{yr}^{-1}]$ (e)						& $6.7$									& $10.1$								& $146$  					  \\
      $\dot{M}_\star^{70~\mu\mathrm{m}}~[M_\odot \mathrm{yr}^{-1}]$ (f) 					& $2.7$									& $7.6$									& $226$	  					  \\
      $\dot{M}_\star^{60~\mu\mathrm{m}}~[M_\odot \mathrm{yr}^{-1}]$ (g)						& $2.0$									& $9.3$									& $281$  					  \\
      $\dot{M}_\star^{24~\mu\mathrm{m}}~[M_\odot \mathrm{yr}^{-1}]$ (h)  					& $2.5$									& $15.3$								& $101$  					  \\
   \hline  \hline  
    \end{tabular}
\end{center}
  \caption{\label{Table_Examples}A comparison of the SFR resulting from our model and other SFR tracers for exemplary galaxies. 
If the input parameters are constrained by observations, we list the reference in the table. In the brackets the typical uncertainty is indicated. Where no measurements are available, we employ the fiducial values as given in Table \ref{Table_Props}. The fluxes at different wavelengths and distances are taken from the NED database. We list the star formation rates $\dot{M}_\star$ resulting from our model and from calibrations in the literature. Notes: 
(a) using equation (\ref{eq_nucrit}), 
(b) full model at a frequency of $\nu=1.4$ GHz, 
(c) using fiducial formulas (\ref{eq_correlationfitMW}) and (\ref{eq_correlationfitM82}), 
(d) full model at a frequency of $\nu=1.5$ GHz, 
(e) calibration (\ref{eq_correlationfitMurphy}),
(f) calibration (\ref{eq_cal70mu}),
(g) calibration (\ref{eq_cal60mu}),
(h) calibration (\ref{eq_cal24mu}),
(i) from private communication with Michael Brown.}
\end{table*}\noindent
\noindent
In Table \ref{Table_Examples} the estimates for the star formation rate of the different galaxies are listed. We present the results for $\dot{M}_\star^\mathrm{1.4~\mathrm{GHz}}$ and $\dot{M}_\star^\mathrm{1.5~\mathrm{GHz}}$ for using our full model, i.e.~calculating the cosmic ray spectrum and the free-free processes with the with the set of observational parameters. In addition we present the result for the SFR when using the observed luminosities and the fiducial formulas as given in (\ref{eq_correlationfitMW}) and (\ref{eq_correlationfitM82}). For M 82, the full model and the fiducial formula result in the same SFR of $6.8~M_\odot\mathrm{yr}^{-1}$, as this galaxy is indeed our fiducial starburst case. We note, however, that the uncertainty in the magnetic field leads to a possible range of SFRs between $2.4$ and $23.1~M_\odot\mathrm{yr}^{-1}$ for M 82. Similarly a range of $\dot{M}_\star=0.4-0.9~M_\odot\mathrm{yr}^{-1}$ is estimated from our model for M 51 at 1.4 GHz. \\
For comparison we determine the SFR also from commonly used tracer, namely the FIR emission which is thermal emission of dust that has been heated via UV radiation from massive stars. We note here that the FIR luminosity is usually defined to be in the range of $8-1000$~$\mu$m. While there is some variation on that in the literature \citep{ChapmanEtAl2000}, the latter uncertainty does not matter too much, as long as the peak emission of starbursts between $10-120$~$\mu$m is included in the observed range. The derivation of the star formation rate from FIR luminosities further depends on the assumed IMF. Assuming a Salpeter IMF \citep{Salpeter1955}, \citet{Kennicutt1998} suggests the following scaling of $\dot{M}_\star$ with the FIR luminosity $L_\mathrm{FIR}$
\begin{eqnarray}
  L_\mathrm{FIR} = 5.79\times10^9~L_\odot~\frac{\dot{M}_\star}{M_\odot \mathrm{yr}^{-1}}.
\label{eq_calFIR}
\end{eqnarray}
We note that the normalization of this relation changes slightly with the IMF. Adopting a \citet{KroupaWeidner2003} IMF, \citet{KennicuttEvans2012} have shown that the normalization decreases by a factor of $0.86$. The FIR luminosity equals roughly 1.7 times the luminosity at $60~\mu\mathrm{m}$ \citep{ChapmanEtAl2000}. Hence, the SFR can be estimated as
\begin{eqnarray}
  \dot{M}_\star^{60~\mu\mathrm{m}} = \frac{60~\mu\mathrm{m}~L_\mathrm{60~\mu\mathrm{m}}}{L_\odot} \frac{M_\odot \mathrm{yr}^{-1}}{1.7\times5.79\times10^9}.
\label{eq_cal60mu}
\end{eqnarray}
This conversion depends in the general case both on the dust temperature as well as on the strength of the interstellar radiation field. In addition we compare to the SFR calibration at $70~\mu\mathrm{m}$ \citep{CalzettiEtAl2010}:
\begin{eqnarray}
  \frac{\dot{M}_\star^{70~\mu\mathrm{m}}}{M_\odot~\mathrm{yr}^{-1}} \approx 2.26\times10^{-10} \frac{c~ L_{70\mu\mathrm{m}}}{70\mu\mathrm{m}~L_\odot}
\label{eq_cal70mu}
\end{eqnarray}
and at $24~\mu\mathrm{m}$ \citep{RiekeEtAl2009}\footnote{We note that the luminosities with an index, e.g.~$L_{70\mu\mathrm{m}}$ and $L_{24\mu\mathrm{m}}$ are spectral luminosities with the units erg/(s Hz). These need to be multiplied with the observing frequencies $c/70\mu\mathrm{m}$ and $c/24\mu\mathrm{m}$, respectively, to gain the physical luminosities with the units erg/s.}:
\begin{eqnarray}
  \frac{\dot{M}_\star^{24~\mu\mathrm{m}}}{M_\odot~\mathrm{yr}^{-1}} \approx 7.84\times10^{-10} \frac{c~ L_{24\mu\mathrm{m}}}{24\mu\mathrm{m}~L_\odot}.
\label{eq_cal24mu}
\end{eqnarray}
Moreover, we list the SFR estimate from the empirical 1.4 GHz calibration suggested by \citet{MurphyEtAl2011} (see equation \ref{eq_correlationfitMurphy}). \\
For the case of M 51 our model yields $\dot{M}_\star^\mathrm{1.4~\mathrm{GHz}}=0.6~M_\odot \mathrm{yr}^{-1}$, both for the full calculation and the fiducial formula. This value is almost a factor three smaller then the one from the $60~\mu\mathrm{m}$ emission, which yields $\dot{M}_\star^\mathrm{60~\mu\mathrm{m}}=2.0~M_\odot \mathrm{yr}^{-1}$. The calibration by \citet{MurphyEtAl2011} results into a higher value of $6.7~M_\odot \mathrm{yr}^{-1}$. Interestingly, if we use the fiducial starburst formula (\ref{eq_correlationfitM82}) for M 51, our model results in a value of $5.4~M_\odot \mathrm{yr}^{-1}$, more comparable to the value from the Murphy calibration. For M 82, which is our fiducial starburst galaxy, our estimates, $\dot{M}_\star^\mathrm{1.4~\mathrm{GHz}}=6.8~M_\odot \mathrm{yr}^{-1}$ and $\dot{M}_\star^\mathrm{1.5~\mathrm{GHz}}=6.4~M_\odot \mathrm{yr}^{-1}$, are slightly smaller than the calibrations from literature. However, the uncertainties of the magnetic field allow for values between $\dot{M}_\star\approx(2-23)~M_\odot \mathrm{yr}^{-1}$. As mentioned before, Arp 220 is not a good candidate for the SFR estimate presented in this paper. Because of the high value of $\nu_\mathrm{crit}$, synchrotron emission is suppressed at 1.4 GHz and thus we expect no correlation with star formation at this frequency. Ignoring this condition and estimating the SFR with the 1.4 GHz flux results into $118~M_\odot \mathrm{yr}^{-1}$ which is about a factor of two smaller than the estimate via the 60 $\mu$m emission and comparable to the one from the 24 $\mu$m emission. The deviations can, however, be larger depending on the contribution from free-free and thermal emission. In general, our physical model for the relation between radio emission and the SFR coincides well with traditional SFR tracers. 


\section{Observations of highly redshifted galaxies}

\subsection{General considerations}
\begin{figure}
  \includegraphics[width=0.45\textwidth]{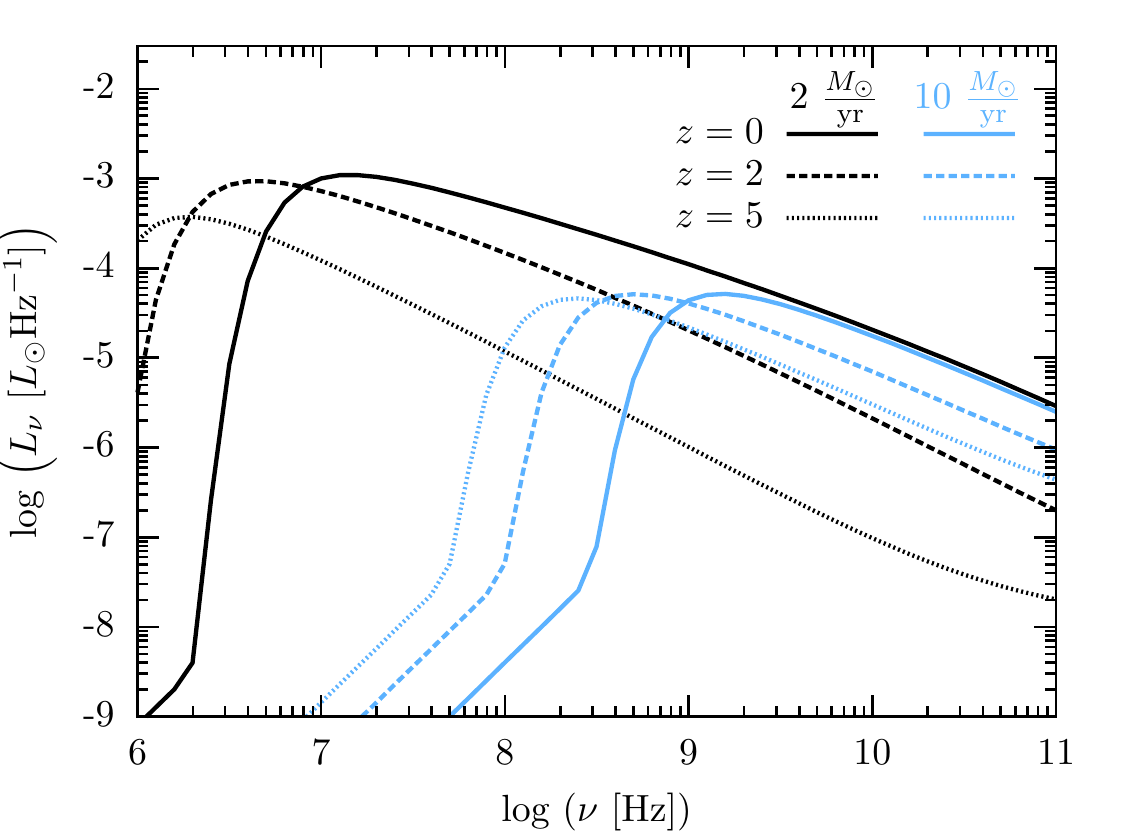}
  \caption{The spectral total radio luminosity $L_\nu$ for our fiducial model of the Milky Way (black lines) and M 82 (blue lines). Different line types correspond to different redshifts $z$.}
\label{plot_Lnu_nu__z}
\end{figure}

\begin{figure}
  \includegraphics[width=0.45\textwidth]{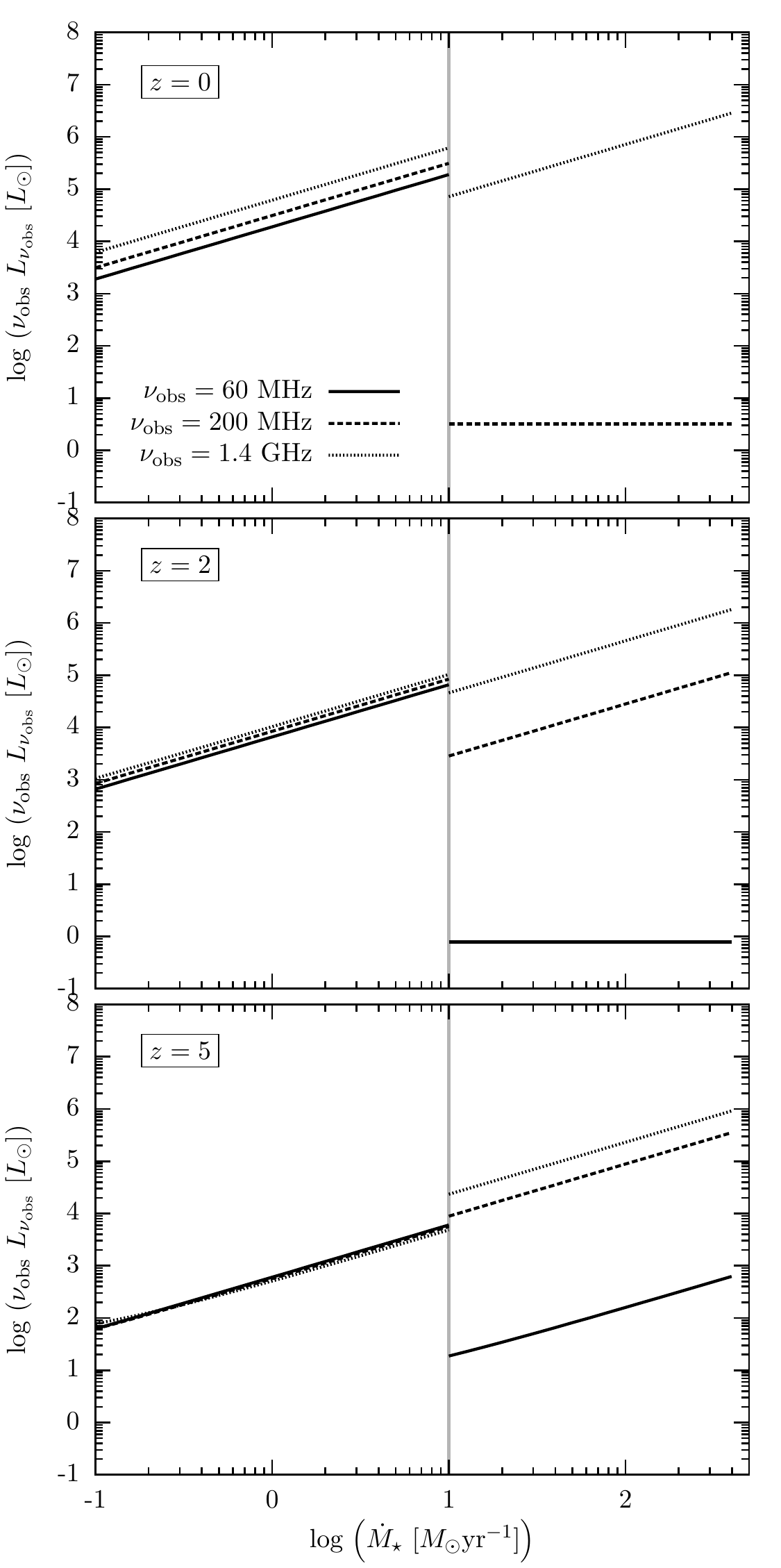}
  \caption{A test of the correlation between radio luminosity $\nu_\mathrm{obs} L_{\nu_\mathrm{obs}}$ and the star formation rate $\dot{M}_\odot$. We explore different observed frequencies and the luminosity at redshift $z=0$ (black lines), $z=2$ (blue lines), and $z=5$ (orange lines). Except for $z$ and $\dot{M}_\odot$ all free parameters used here are for the fiducial galaxies. We note that the line at 10 $M_\odot\mathrm{yr}^{-1}$ is only for illustrative purpose and does not imply a physical transition.}
\label{plot_L_SFR__z3}
\end{figure}

An important goal of the upcoming radio telescopes is to study galaxies at early times and their evolution to the present day. Theory predicts young galaxies to be smaller and denser and to have higher star formation rates than their local counterparts. Observations at high redshifts are crucial to constrain theories of evolution of galaxies and galactic star formation. \\
Hypothetically, when moving a galaxy of fixed density and star formation rate to higher redshift, one expects two effects on the resulting radio luminosity. First, the number of cosmic rays is reduced, as they lose energy faster via inverse Compton scattering with the stronger CMB \citep{Murphy2009,LackiEtAl2010b,SchleicherBeck2013,SchoberEtAl2016}. This results in less synchrotron emission and thus needs to be taken into account when estimating the SFR. Second, spectral signatures, like the critical frequency, move to smaller $\nu$ in the observed frame. In fact the critical frequency in the observed frame is
\begin{eqnarray}
  \frac{\nu_\mathrm{crit,obs}}{10^9~\mathrm{Hz}} & = &  \frac{1}{1+z}\left(0.082\left(\frac{T_\mathrm{e}}{\mathrm{K}}\right)^{-1.35} \left(\frac{EM(n)}{\mathrm{cm^{-6}~pc}}\right) \right)^{1/2.1}.  \nonumber \\
\label{eq_nucritobs}
\end{eqnarray}
Consequently, non-thermal radio emission from highly redshifted galaxies can be used also at lower frequencies. \\
\begin{figure}
  \includegraphics[width=0.45\textwidth]{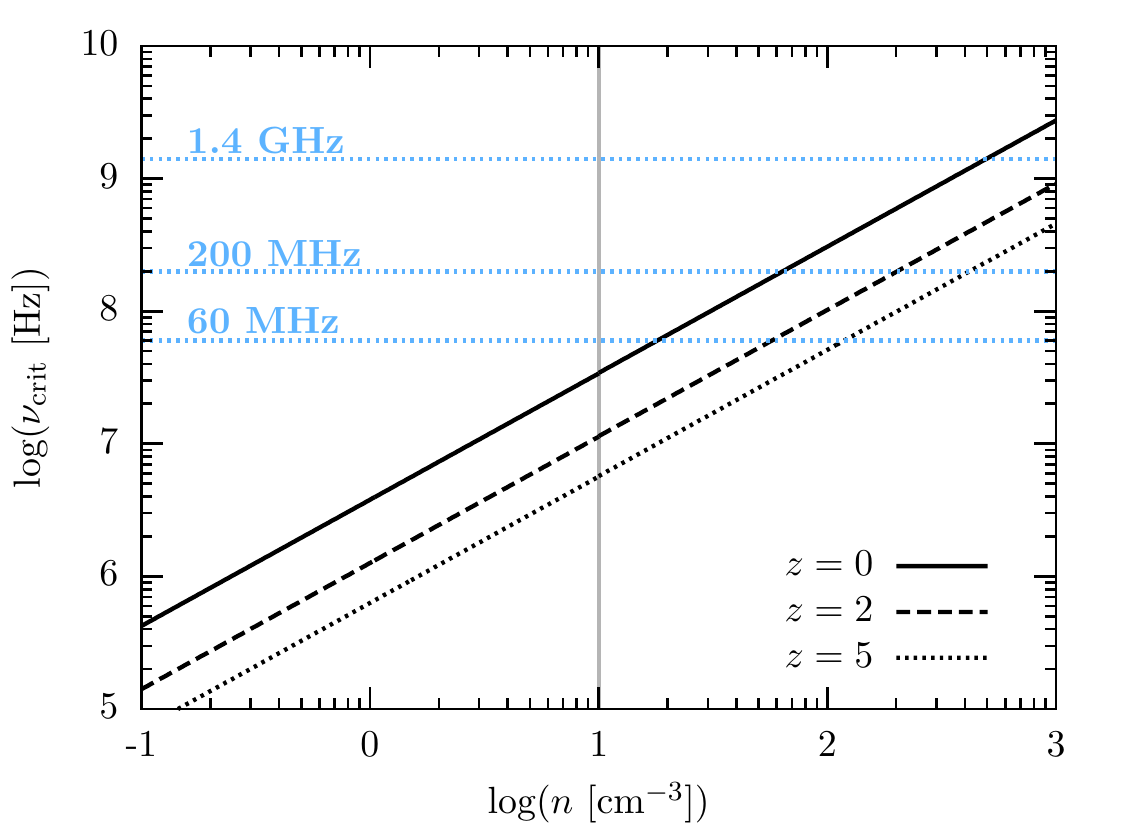}
  \caption{The critical frequency $\nu_\mathrm{crit}$ below which the optical depth $\tau_\mathrm{ff}<1$ as a function of the gas density $n$. We present the result for the fiducial galaxies at different redshifts $z$.}
\label{plot_nucrit_n__z}
\end{figure} \noindent
Figure \ref{plot_Lnu_nu__z} shows how the spectra change with increasing redshift for the Milky Way model (black lines) and the M 82 model (blue lines). The critical frequency, i.e.~the turnover of the spectrum, shifts to smaller $\nu$ as $z$ increases. For our fiducial starburst galaxy non-thermal radio emission cannot be employed as a SFR tracer at 60 MHz and $z=0$, while it becomes possible again at the same $\nu$ and $z=5$. For the starburst case, the normalization of the spectrum is not affected significantly by higher redshift, as the intrinsic radiation is very strong compared to the CMB, also at $z=5$. Similarly, Figure \ref{plot_nucrit_n__z} presents how $\nu_\mathrm{crit}$ decreases with increasing redshift. Hence, a galaxy which is no candidate for our method in the local Universe, can show a correlation between the SFR and the non-thermal radio emission if it was at higher redshift. This is illustrated in Figure \ref{plot_L_SFR__z3} where we show how the correlation of radio luminosity at different fixed observed frequencies is reestablished as $z$ increases. We note here, that the total radio luminosity decreases with $z$, as the number of cosmic rays is reduced by stronger inverse Compton losses. This effect is less important for starbursts, as they host a very strong intrinsic interstellar radiation field, which is much stronger than the CMB up to high redshifts.   \\
\noindent

\subsection{Exemplary galaxies at high redshift}
Detailed observations of highly redshifted galaxies are possible if they are gravitationally lensed by massive foreground galaxy clusters. For example, \citet{IvisonEtAl2010} estimate the properties of SMM J2135-0102, also known as the cosmic eyelash, which has a redshift of $z=2.3$. They report a density of $n=10^3~\mathrm{cm}^{-3}$. Assuming a scale height of $100$ pc, which is based on an estimate of the galaxy's star forming regions \citep{SwinbankEtAl2010}, an electron temperature of $T_\mathrm{e}=5000$ K, $f_\mathrm{ion}=0.1$, and $f_\mathrm{fill}=0.2$, we find a critical frequency of $\nu_\mathrm{crit}=5.94\times10^8$ Hz. Inserting a spectral radio luminosity of $L_\mathrm{1.4~GHz}=9\times10^{23}~\mathrm{W~Hz}^{-1}$ \citep{IvisonEtAl2010} into the fiducial formula (\ref{eq_correlationfitM82}) yields a SFR of 456 $M_\odot\mathrm{yr}^{-1}$. This is comparable to the estimate of 400 $M_\odot\mathrm{yr}^{-1}$ based on the intrinsic rest-frame 8-1000 $\mu$m luminosity \citep{IvisonEtAl2010}. \\
\citet{SmolcicEtAl2012} identify three high-$z$ submillimeter galaxies with radio and infrared counterparts, namely Cosbo-3, Cosbo-8, and AzTEC/C1. For Cosbo-3 they report a 20 cm luminosity of $L_{20~\mathrm{cm}}\approx 2\times10^{24}~\mathrm{W~Hz}^{-1}$. While no information about the parameters for estimating the critical frequency of the source is available, we use our fiducial starburst formula (\ref{eq_correlationfitM82}) with which we find a SFR of approximately 1000 $M_\odot\mathrm{yr}^{-1}$, while the Murphy calibration (\ref{eq_correlationfitMurphy}) yields 1300 $M_\odot\mathrm{yr}^{-1}$. Using equation (\ref{eq_calFIR}) we find from the infrared luminosity of $1.5\times10^{13}~L_\odot$ a SFR of 2600 $M_\odot\mathrm{yr}^{-1}$. For Cosbo-8, \citet{SmolcicEtAl2012} report $L_{20~\mathrm{cm}}\approx 10^{25}~\mathrm{W~Hz}^{-1}$, which translates in a SFR estimate of 5100 $M_\odot\mathrm{yr}^{-1}$ using equation (\ref{eq_correlationfitM82}). In contrast, the infrared luminosity of $1.1\times10^{13}~L_\odot$ implies $\dot{M}_\star\approx1900~M_\odot\mathrm{yr}^{-1}$. For the third galaxy discussed in \citet{SmolcicEtAl2012}, AzTEC/C1, only a radio luminosity is listed. With the value of $L_{20~\mathrm{cm}}\approx (6-10)\times10^{24}~\mathrm{W~Hz}^{-1}$, we estimate a star formation rate of $(3000- 5100) M_\odot\mathrm{yr}^{-1}$. \\
In general, as the detailed properties of these galaxies are unknown, we consider the above SFR estimates based on the radio and FIR luminosities in rather good agreement. We note, that for Cosbo-3 and SMM J2135-0102 the calibrations in both wavelengths yield comparable SFRs, while the radio-based SFR estimate for Cosbo-8 is enhanced as compared to the IR-based one. This might be an indication of the presence of a weak active galactic nuclei (AGN), consistent with the X-ray luminosity of $6.8\times10^{43}~\mathrm{erg~s}^{-1}$ \citep{SmolcicEtAl2012}. An alternative explanation of the enhanced radio-based SFR would be a rather strong magnetic field.

\subsection{Uncertainties due to galaxy evolution and the presence of AGNs}
\label{sec_highz-uncertainties}
\begin{figure}
  \includegraphics[width=0.45\textwidth]{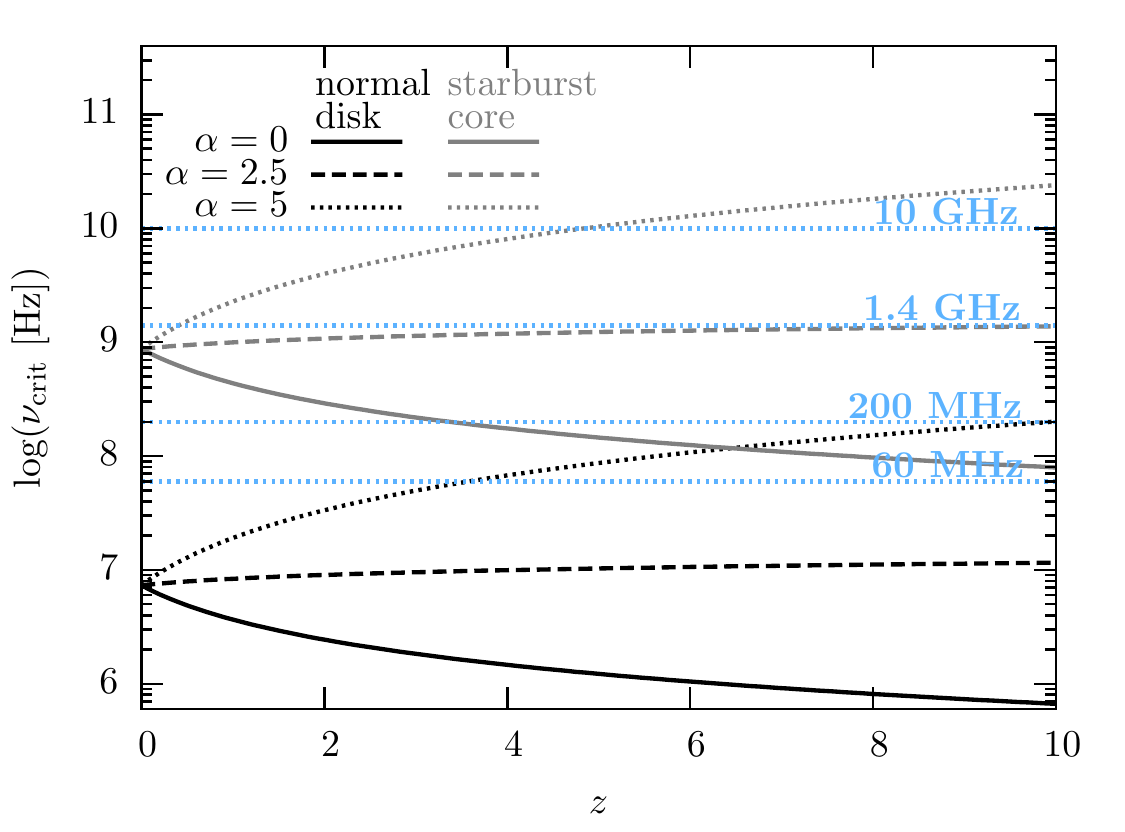}\hfill
  \caption{The change of the observed critical frequency $\nu_\mathrm{crit}$ with redshift $z$ for different galaxy evolution models. The redshift dependencies are absorbed in a scaling of the emission measure $EM \propto (1+z)^{\alpha}$, where $\alpha$ is explored between 0 and 5. Black curves refer to the case of our fiducial disk model, while the gray curves refer to the starburst model.}
\label{plot_nucrit_z}
\end{figure}

Galaxies at high redshift seem to have different properties than their local counterparts. Observations suggest that they are more compact \citep{DaddiEtAl2005,WilliamsEtAl2014,BelliEtAl2014} and possibly have a modified ISRF \citep{BetherminEtAl2015}. These properties could enter our model for the non-thermal radio luminosity and affect the estimate of the star formation rate. While we are not assuming a specific galaxy evolution model in this paper, but rather provide an independent formalism based on given galaxy properties, we briefly estimate the effect of redshift evolution in this section. \\
Semi-analytical calculations result in a redshift evolution of the galactic radius $R$ proportional to $(z+1)^{-1}$ if the galaxy's mass is constant \citep{MoMaoWhite1998}. This scaling is supported by observations of \citet{OeschEtAl2010} who find a scaling of the half-light radius with $(1+z)^{-1.12 \pm 0.17}$. Assuming a similar redshift scaling of the galactic scale height, the gas density evolves as $n(z)\propto R(z)^{-2} H(z)^{-1} \propto (1+z)^3$. As a result, the emission measure (\ref{eq_EM}) would scale as $EM \propto n(z)^2 H(z) \propto (1+z)^5$, which can have dramatic consequences for the critical frequency (\ref{eq_nucritobs}). To explore this effect we use a general scaling of
\begin{equation}
  EM \propto (1+z)^{\alpha}
\label{eq_EMz}
\end{equation}
with a range from $\alpha=0$ for a negligible galaxy evolution up to $\alpha=5$ for a strong evolution with redshift. We note, however, that the free parameter $\alpha$ as introduced in equation (\ref{eq_EMz}) is difficult to constrain quantitatively. In fact, as $EM$ and hence also $\nu_\mathrm{crit}$ are degenerated in various other quantities besides $n$ and $H$, the redshift evolution might be less significant then $\alpha=5$. The emission measure (\ref{eq_EM}) also depends on the ionization degree and the filling factor, which both could also vary in young galaxies. Especially, the ionization degree can play an important role, as $\nu_\mathrm{crit}$ scales almost linear with $f_\mathrm{ion}$, exactly as with $n$. Additionally, an evolution of the electron temperature enters the critical frequency, as $\nu_\mathrm{crit}\propto T_\mathrm{e}^{-0.64}$. \\
For generic evolution scenarios as given in (\ref{eq_EMz}), we present the observed critical frequencies as a function of redshift for our fiducial models in Figure \ref{plot_nucrit_z}. While $\nu_\mathrm{crit}$ decreases with $z$ for a galaxies with constant densities and scale heights, i.e.~$\alpha=0$, it stays roughly constant for $\alpha=2.5$, and even increases for a strong galaxy evolution with $\alpha=5$. In the latter case, non-thermal radio emission can only be used as a SFR tracer for high observing frequencies. In fact, for $z\gtrsim4.5$ and our fiducial starburst model, flux measurements below 10 GHz can only be used for estimating an upper limit of the SFR. A hint towards a less strong evolution of $\nu_\mathrm{crit}$ with $z$ is the observation of the FIR-radio correlation up to $z\approx3$ \citep{MagnelliEtAl2015,PannellaEtAl2015}. This observational fact is consistent with a value of $\alpha=2.5$ in our fiducial starburst model.   \\
An additional potential caveat for using radio emission to study star formation at high redshift, is that in future deep surveys also a large number of AGNs will be seen. They can dominate the radio regime and will make it impossible to estimate the star formation rate of their host galaxies. Galaxies hosting an AGN can be excluded from the data set by identifying them with X-ray observations \citep{TreisterEtAl2009}, unless they X-ray emission is absorbed by a Compton-thick envelope, or by detecting radio jets if the latter can be spatially resolved. On the other hand, if the observed radio luminosity is higher than the one of the average population, i.e.~if the source lies significantly above the FIR-radio correlation, it should be regarded as an AGN. In this case our SFR estimate as given in (\ref{eq_correlationfitM82}) only provides an upper limit.  \\


\section{Conclusion}

In this paper we study the conditions under which non-thermal radio emission can be used to estimate the star formation rate. The underlying physical connection between the two quantities are cosmic rays, which are produced in supernova shock fronts and thus are linked to a galaxy's SFR. These high-energy charged particles emit synchrotron radiation when traveling through the magnetized interstellar medium. Here we have derived an estimate of the galactic radio luminosity as a function of the SFR, and carefully discussed the frequency regime where such correlations are applicable. \\
The main results of this work are:
\begin{itemize}
\item{Synchrotron emission is proportional to the star formation rate. The main dependency results from the production of cosmic rays which takes place in supernova shock fronts and is thus related to the supernova rate, which in turn scales as the SFR.}
\item{If the gas density is too high, however, synchrotron emission is absorbed by the free-free process. This absorption is especially important at low frequencies. In fact, below a critical frequency $\nu_\mathrm{crit}$ (see equation \ref{eq_nucritobs}) radio emission cannot be employed for measuring SFRs at very low frequencies. Besides the gas density, the value of $\nu_\mathrm{crit}$ depends on the ionization degree, the scale height, the electron temperature, and the filling factor.} 
\item{At high redshifts, the observed radio spectrum and with it the critical frequency moves to lower frequencies (see equation \ref{eq_nucritobs}). Here radio emission can again be used to determine SFRs also for dense young galaxies.}
\item{The general relation between SFR and radio luminosity at different frequencies is described by equation (\ref{eq_Lnu}). In addition, we provide simple fiducial formulas for normal disk galaxies and starburst systems in equations (\ref{eq_correlationfitMW}) and (\ref{eq_correlationfitM82}). At high $z$ these relations should be used with caution, as they depend on galaxy evolution scenarios (see discussion in section \ref{sec_highz-uncertainties}).}
\end{itemize}
We have applied our method to local exemplary test galaxies which are presented in Table \ref{Table_Examples}. The SFRs determined from non-thermal radio emission are comparable the ones resulting from the FIR fluxes. \\
Our method should be a useful tool for future deep radio surveys, for example with the SKA. Most surveys will be performed at a fixed observing frequency. Here, especially without the information from the spectral flux distribution, caution is required. First of all, the critical frequency $\nu_\mathrm{crit}$ should be estimated, as radio-SFR calibrations like the one presented here should only be applied if the observing frequency is above $\nu_\mathrm{crit}$. Otherwise the synchrotron emission is absorbed and the radio flux is not correlated with the SFR. Only if synchrotron radiation dominates the spectrum at the observed frequency, the SFR can be estimated with the fiducial formulas given in equations (\ref{eq_correlationfitMW}) for normal disk galaxies and in equation (\ref{eq_correlationfitM82}) for starbursts. We note however, that our model includes several free parameters which are summarized in Table \ref{Table_Props}. The accuracy of the SFR estimate can be increased if additional parameters of the galaxy, like the gas density, the ionization degree, and the scale height, are known. Additionally, the thermal and the free-free contribution might be important especially at higher radio frequencies, or at even at lower frequencies if the spectrum is highly redshifted. The latter contributions depend on the galaxy's temperature and surface area. For example, the thermal peak in the spectrum of the starburst galaxy Arp 220 occurs above approximately $100$ GHz (see Figure \ref{plot_RealSpectra}). In the observed spectrum of a similar galaxy at $z=10$, the thermal bump would be located at around $9$ GHz. Additionally, the presence of an AGN should be excluded, which could easily dominate the galactic radio emission.

\section*{Acknowledgements}
We are grateful to the referee for carefully reading our manuscript and for his or her highly insightful comments. This work has been financially supported by {\em Nordita} which is funded by the Nordic Council of Ministers, the Swedish Research Council, and the two host universities, the {\em Royal Institute of Technology} (KTH) and {\em Stockholm University}. DRGS thanks for funding through Fondecyt regular (project code 1161247), through the ``Concurso Proyectos Internacionales de Investigaci\'on, Convocatoria 2015'' (project code PII20150171), and from the Chilean BASAL Centro de Excelencia en Astrof\'isica y Tecnolog\'as Afines (CATA) grant PFB-06/2007. We further thank for funding through the {\em Deutsche Forschungsgemeinschaft} (DFG) in the {\em Schwer\-punkt\-programm} SPP 1573 ``Physics of the Interstellar Medium'' under grants  BO 4113/1-2, KL 1358/18.1, KL 1358/19.2, SCHL 1964/1-1, and SCHL 1964/1-2. In addition we thank the DFG for support via the SFB 881 ``The Milky Way System'' in the sub-projects B1 and B2. Also we acknowledge financial support by the {\em European Research Council} under the {\em European Community's Seventh Framework Programme} (FP7/2007-2013) via the ERC {\em Advanced Grant} STARLIGHT (project number 339177).







\bsp	
\label{lastpage}
\end{document}